\documentclass{emulateapj}
\usepackage{amsmath}
\newcommand{\beq}{\begin{equation}}
\newcommand{\eeq}{\end{equation}}

\newcommand{\obj}{HM Cnc}

\newcommand{\rvcorr}{$390\pm40$\,km\,s$^{-1}$}
\newcommand{\rvcorrHeII}{$260\pm40$\,km\,s$^{-1}$}
\newcommand{\rvshiftHeII}{$650\pm50$\,km\,s$^{-1}$}
\newcommand{\qlow}{$q\geq0.33\pm0.06$}
\newcommand{\qhigh}{$q\leq0.67\pm0.12$}
\newcommand{\q}{$q=0.50\pm0.13$}

\newcommand{\ions}[2]{\mbox{#1\,\textsc{#2}}}

\begin{document}

\shorttitle{A 5.4-minute orbital period in HM Cnc}

\title{Spectroscopic Evidence for a 5.4-Minute Orbital Period in HM Cancri}

\shortauthors{Roelofs, Rau, et al.}

\author{Gijs H.\,A. Roelofs}
\affil{Harvard--Smithsonian Center for Astrophysics, 60 Garden Street, Cambridge, MA 02138, USA;\\groelofs@cfa.harvard.edu}

\author{Arne Rau}
\affil{Astronomy Department, California Institute of Technology, Pasadena, CA 91125, USA}
\affil{Max-Planck Institute for Extra-Terrestrial Physics, Giessenbachstr.\ 1, 85748 Garching, Germany;\\arau@mpe.mpg.de}

\author{Tom R. Marsh, Danny Steeghs}
\affil{Department of Physics, University of Warwick, Coventry, CV4 7AL, United Kingdom}

\author{Paul J. Groot, Gijs Nelemans}
\affil{Department of Astrophysics, Radboud University Nijmegen, PO Box 9010, 6500 GL, Nijmegen, The Netherlands}

\begin{abstract}
HM Cancri is a candidate ultracompact binary white dwarf with an apparent orbital period of only 5.4 minutes, as suggested by X-ray and optical light-curve modulations on that period, and by the absence of longer-period variability.
In this Letter we present Keck-I spectroscopy which shows clear modulation of 
the helium emission lines in both radial velocity and amplitude on the 5.4-minute 
period and no other.
The data strongly suggest that the binary is emitting \ions{He}{i} 4471 from the irradiated face of the cooler, less massive star, and \ions{He}{ii} 4686 from a ring around the more massive star. From their relative radial velocities, we measure a mass ratio \q.
We conclude that the observed 5.4-minute period almost certainly represents the orbital period of an interacting binary white dwarf. We thus confirm that \obj\ is the shortest-period binary star known: a unique test for stellar evolution theory, and one of the strongest known sources of gravitational waves for the \emph{Laser Interferometer Space Antenna (LISA)}.
\end{abstract}

\section{Introduction}

Two interacting binary stars have been discovered which appear to have orbital periods shorter than ten minutes: V407 Vul, with a period of 9.5 minutes \citep{Mot96,Cro98}, and HM Cnc, with a period of 5.4 minutes \citep{Isr02}. The uniquely short period of 5.4 minutes, if it is the orbital period, implies that \obj\ must have formed from two white dwarfs, driven together as a result of gravitational-wave radiation. It may currently be experiencing stable mass transfer through Roche-lobe overflow. Because it is potentially so extreme and unique, substantial effort has been put into unveiling \obj's true nature, but as yet without conclusive results.

The key observational data show that: (1) there is no evidence for variability on periods other than 5.4 and 9.5 minutes in \obj\ and V407 Vul, respectively \citep{Ram00,Ram02a}; (2) the optical flux maxima lead the X-ray maxima by about 90 degrees in both systems \citep{Bar07}; and (3) the periods are decreasing in both systems \citep{Str02,Str03,Str04,Str05,Hak03,Hak04,Isr04}.

Three competing models have been proposed for V407 Vul and \obj. One of them, the `Intermediate Polar (IP)' model, predicts that these systems are not in fact ultracompact binaries but have rather mundane orbital periods of several hours. The ultrashort periods then represent the spins of magnetic white dwarfs \citep{Nor04}. The X-rays as well as the variable optical flux originate from the accretion flow crashing onto the magnetic poles of the magnetic white dwarf during part of the accretor's spin cycle. The spin-up of the X-ray and optical periods may be expected since the magnetic white dwarf is accreting matter of high specific angular momentum. The absence of variability on the much longer orbital period could be explained if the orbital planes of both systems are viewed exactly face-on. The observed phase offsets between the optical and X-ray light-curves are difficult to explain \citep{Nor04}, and the emission lines are unusually weak for IPs; in V407 Vul there appear to be no emission lines at all \citep{Ram02b,Ste06}.

The second, `Unipolar Inductor (UI)' model, is essentially a more energetic version of the Jupiter--Io system \citep{Wu02,Dal06,Dal07}. A magnetic white dwarf is orbited by another, non-magnetic one; the magnetic field induces an electrical potential across the non-magnetic white dwarf which sets up currents along magnetic flux tubes connecting the two stars. Ohmic dissipation of these currents in the flux tubes' footpoints on the magnetic star gives rise to the X-rays. In this model the 5.4-minute period is the orbital period, but the two stars are detached. Since the binary loses angular momentum due to gravitational-wave radiation, it is expected to evolve towards shorter orbital periods, as observed. The main problem is the offset between the optical and X-ray flux, which requires the footpoints on the magnetic star's surface to be almost 90 degrees ahead of the orbiting non-magnetic star in both systems \citep{Bar05,Bar07}.

The third, `AM CVn' model, has a Roche-lobe filling white dwarf losing mass to a more massive white dwarf (\citealt{Mar02,Ram02b}, or \citealt{Cro98} for a magnetic version). The orbital period is 5.4 minutes in the case of \obj, and the accretion stream hits the accretor directly without forming a disk. The phase offset between optical and X-ray flux is naturally accounted for, since the accretion stream will deflect and hit the accretor off-axis. The absence of longer periods is also expected, but the observed decrease of the periods in \obj\ and V407 Vul is considered problematic for this model \citep{Str02,Str04}, although solutions to this problem have been put forward (\citealt{Del06,DAn06}; Roelofs \& Deloye in preparation).

This Letter describes our search for kinematic evidence, using phase-resolved spectroscopy of \obj, for or against these different models. Section \S\,\ref{sec:obs} describes our observations and data reduction. We present our results in \S\,\ref{sec:results} and discuss their implications in \S\,\ref{sec:conclusion}.

\section{Observations and Data Reduction}
\label{sec:obs}

\begin{table*}
\begin{center}
\caption{Log of our Keck/LRIS observations of HM Cnc.}
\label{table:log}
\begin{tabular}{l l l l l l}
\hline
Date		&UT		&Spectra	&Weather						&Total S/N\\
			&		&(60\,s)	&								&(at 4200\,\AA)\\
\hline
\hline
2009/01/26	&08:10--12:30		&136		&1.1$''$, absorption early		&78\\
2009/01/27	&07:24--13:14		&223		&1.1$''$, clear					&115\\
2009/03/31	&05:52--06:50		&41			&1.3$''$, clear					&41\\
\hline
\end{tabular}
\end{center}
\end{table*}

After unsuccessful attempts in 2005, 2006 \& 2007 to obtain phase-resolved spectroscopy of \obj, we finally met acceptable weather conditions at Keck-I on January 25 \& 26 and March 30 of 2009. We took spectra using a 1.5$''$ slit and the 600 grooves/mm grism on the blue side of LRIS, the Low-Resolution Imaging Spectrograph \citep{Oke95}, for an effective spectral resolution of about 300 km/s. In order to resolve the 321.5-second period in \obj, we limited the exposure times to 60 seconds. The blue CCD, with a plate scale of 0.135$''$/pixel, was binned $4\times 4$ pixels to reduce the read-out noise while still sampling resolution elements with approximately 2 (binned) pixels. The binning furthermore reduced the dead-time between exposures to 27 seconds.

Data reduction was done using standard \textsc{iraf} tasks. To optimally extract the very faint individual spectra (\obj\ has optical magnitudes $U\approx 19.6, B\approx 20.7$; \citealt{Isr02}), reference aperture profiles were created from averaged blocks of typically $\sim$30 successive frames to increase the photon statistics and remove cosmic rays from the profile. Spectra were grouped together based on the requirement that spatial center and profile (seeing) on the CCD be (near-)identical. Wavelength calibration was achieved using Hg\-Cd\-Zn\-Ne\-Ar arc exposures. A fit of 30 arc lines left 0.35\AA\ root-mean-square residuals. No detectable arc drift occurred during the night, and the drift between the two nights was well below one pixel. We thus applied a common dispersion solution to all spectra, only applying a small shift to the first night's spectra to tie the strong [\ions{O}{i}] 5577 sky emission lines together. The relative wavelength calibration of all spectra is estimated to be better than 0.1\AA.

The time stamp and velocity scale of each spectrum were transformed to the solar system's barycenter. Time stamps were checked several times during our run using Universal Time clocks available online and estimated to be correct to one second (or better). The spectrophotometric standard star Feige 34 was used for calibrating the instrumental response; variable transparency and seeing precluded an absolute flux calibration. Table \ref{table:log} summarizes our observing log.

\section{Results}
\label{sec:results}

\subsection{Average Spectrum}

\begin{figure}
\begin{center}
\includegraphics[width=84mm]{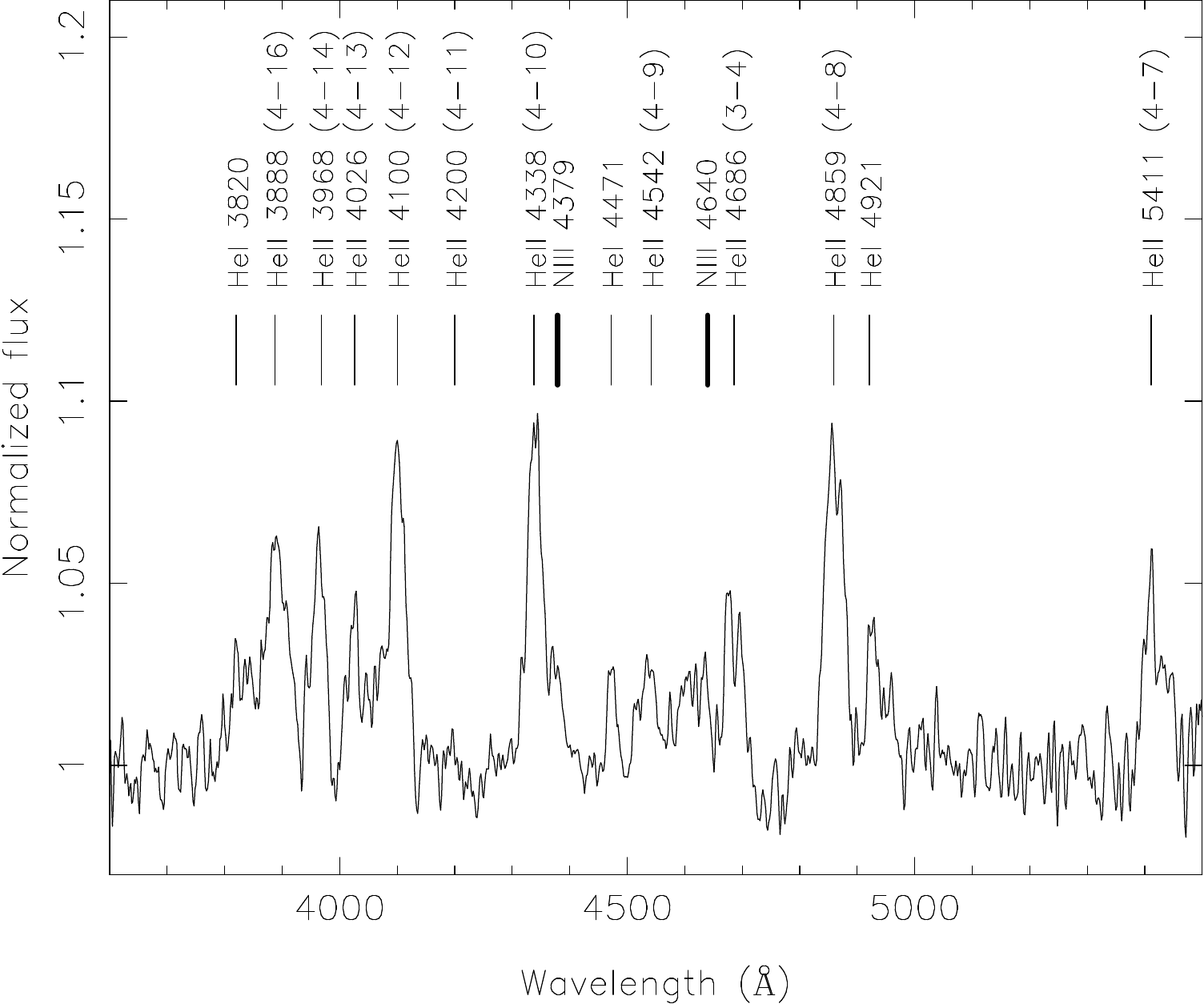}
\caption{Average Keck-I/LRIS spectrum of \obj, normalized to the continuum. The most obvious spectral lines are labeled. The \mbox{He\,\textsc{ii}} 3888 and 4026 lines may contain a contribution from coincident \mbox{He\,\textsc{i}} lines.}
\label{fig:spectrum}
\end{center}
\end{figure}

The average Keck spectrum of \obj\ is shown in Fig.\ \ref{fig:spectrum}. As previously reported \citep{Isr02} the spectrum is dominated by ionized helium emission lines. Their full-width at half-maximum (FWHM) of $\sim$2500\,km\,s$^{-1}$ is well resolved by the 300\,km\,s$^{-1}$ resolution of our spectra. Lines of neutral helium are also present but much weaker. Our higher-quality spectrum confirms earlier findings that the even-term transitions of the \ions{He}{ii} Pickering series are stronger than the odd-term ones (see Fig.\ \ref{fig:spectrum}), which was interpreted by \citet{Nor04} and \citet{Rei07} as evidence for the presence of hydrogen. In particular the absence of \ions{He}{ii} 4200 in our spectrum is striking, and most easily explained if one assumes that there is indeed hydrogen in \obj\ (for simplicity, we will refer to \ions{He}{ii} Pickering lines in the remainder of this paper).

In addition to the \ions{N}{iii} 4640 line in the Bowen blend identified by \citet{Isr02}, the strong \ions{N}{iii} 4379 line appears to be present to the red of \ions{He}{ii} 4338. Apparent emission lines near 4600\,\AA\ and 4800\,\AA\ could be due to \ions{N}{iv}, while a faint feature to the red of \ions{He}{i} 4471 may be \ions{N}{iii} 4514. The presence of several nitrogen lines appears to be at odds with the suggestion (based on fits to X-ray spectra) that \obj\ may have an unusual chemical abundance pattern and be underabundant in nitrogen in particular \citep{Str08}. We note that nitrogen-rich matter would be expected due to the CNO-cycle if \obj\ is an ultracompact binary, unless the material experienced helium burning \citep{Yun08}, and that nitrogen indeed appears to be abundant in most AM CVn stars (see \citealt{Roe09}).

\begin{figure*}
\begin{center}
\includegraphics[width=\textwidth]{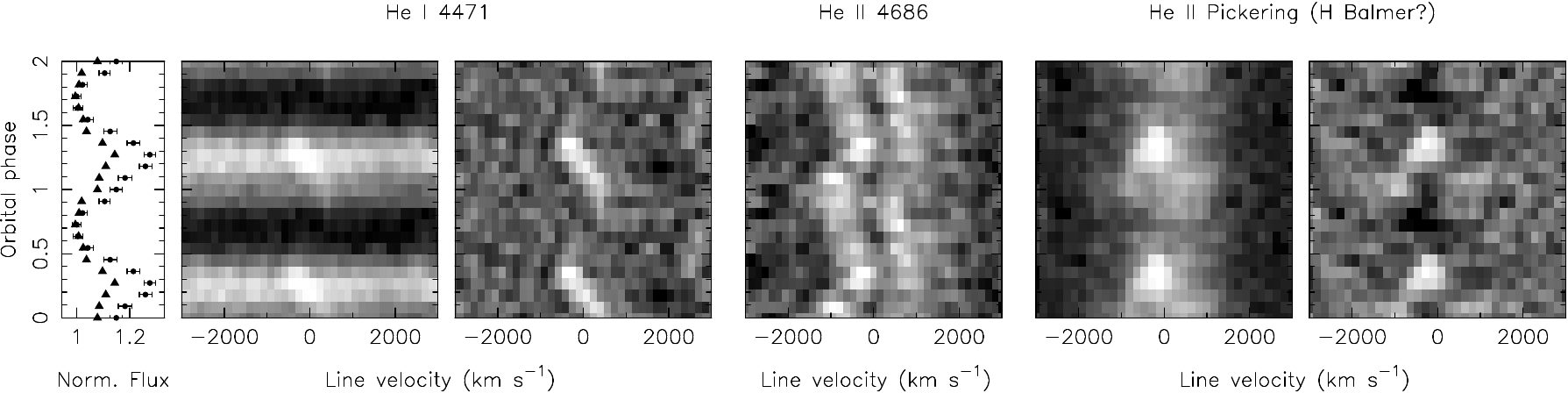}
\caption{Trailed spectra of \obj, folded on the ephemeris of Eq.~(\ref{eq:ephemeris}), and resampled into 11 phase bins. The left-most panel shows the continuum flux variations near the \ions{He}{i} 4471 line (dots with error bars) as well as the line emission intensities, measured between $\pm$800 km/s relative to the continuum (triangles; scaled up by a factor five). In each but the left-most trailed spectrum the varying continuum flux has been subtracted to highlight the emission line variability. The middle trail shows the double-peaked \ions{He}{ii} 4686 line moving in anti-phase with the \ions{He}{i} 4471 line. The right two panels show the sum of the three strongest \ions{He}{ii} Pickering series lines (4100, 4338 \& 4859\,\AA; see Fig.\ \ref{fig:spectrum}), where in the right-most panel the average spectrum has been subtracted to highlight the time-varying component. The grey-scales are linear between the dimmest and brightest pixels, where the brightest pixels have the highest flux densities.}
\label{fig:trails}
\end{center}
\end{figure*}

\subsection{Ephemeris and Period Derivative}

When we phase-fold our spectra using the ephemeris of \citet{Bar07}, we observe continuum flux variations that lag slightly ($\sim$30 degrees in phase) behind \citet{Bar07}'s photometry. This is not unexpected given, in particular, the uncertainty in the period derivative measured by \citet{Bar07}, which translates into a phase uncertainty of about this magnitude.

By matching the phases of the variable $g$-band flux in our time-resolved spectra with the $g$-band photometry of \citet{Bar07}, we can obtain a refined ephemeris for HM Cnc. This will not be a very precise refinement since our spectra only marginally sample the orbital period, and since atmospheric transparency and seeing were variable. However, because we have hundreds of spectra and the folding period is much shorter than the typical timescales for sky quality variations, we can still resolve this shift. The easiest way to correct the phase shift is to reduce the frequency $\nu$ and its time derivative from \citet{Bar07} by approximately 1.3\,$\sigma$:
\begin{align}
t_0\,\mathrm{(barycentric\ TDB)} &\equiv 53 009.889 943 753\nonumber\\
\nu\,\mathrm{(Hz)}               &= 0.003 110 138 11(10)\nonumber\\
d\nu/dt\,\mathrm{(Hz\ s^{-1})}   &= 3.57(2)\times 10^{-16}\label{eq:ephemeris}
\end{align}
where numbers in parentheses are the uncertainties in the corresponding number of last decimals. The new error on the frequency derivative is estimated as the error that gives the same phase uncertainty (at our epoch) as the error on the frequency. The error on the frequency has been kept the same. This ephemeris is used throughout this paper.

\subsection{Time-Resolved Spectrum}

The left panels of Fig.\ \ref{fig:trails} show the time-resolved (trailed) spectrum of \obj\ around the \ions{He}{i} 4471 line, folded on the above ephemeris (eq.\ (\ref{eq:ephemeris})). The continuum flux is seen to follow broadly the same pattern as in \citet{Bar07} after aligning the phases: during a cycle the continuum flux increases slowly and decreases more rapidly. The scatter in our continuum fluxes, shown in the left-most panel of Fig.~\ref{fig:trails}, is due to differences in the average sky quality between phase bins rather than due to noise in individual spectra.

Clear spectral line variations are seen in the \ions{He}{i} 4471 line, in the form of an `S-wave' feature that is visible for about half the period while it moves from red-shifted to blue-shifted wavelengths. The intensity of the S-wave in the \ions{He}{i} 4471 line follows the intensity of the variable continuum flux, suggesting that they originate from the same region. In the remaining panels of Fig.\ \ref{fig:trails} the intrinsically variable continuum flux has been fitted and subtracted to highlight the line variability. Here the \ions{He}{ii} 4686 line is plotted as well as the sum of the three strongest \ions{He}{ii} Pickering series lines (4100, 4338 \& 4859\,\AA; see Fig.\ \ref{fig:spectrum}). The \ions{He}{ii} lines behave quite differently from the \ions{He}{i} 4471 line. The \ions{He}{ii} 4686 line is double-peaked and appears to wobble in anti-phase with respect to the \ions{He}{i} 4471 line. The \ions{He}{ii} Pickering lines have a strongly modulated, narrow component which may be moving from blue-shifted to red-shifted in phase with (and at the same radial velocity of) the \ions{He}{ii} 4686 line. In addition the \ions{He}{ii} Pickering lines have a fairly constant, broad component with a FWHM of $\sim$2500\,km\,s$^{-1}$, which matches the FWHM of the (double-peaked) \ions{He}{ii} 4686 profile.

We measure, from a linear back-projection Doppler tomogram \citep{Mar88}, a radial velocity semi-amplitude of \rvcorr\ for the S-wave feature in the \ions{He}{i} 4471 line (corrected for a $\sim$5\% bias due to finite exposure times, which we modeled using synthetic data). The error was estimated from a large ensemble of Doppler tomograms calculated from bootstrap samples of our 400 spectra. For the double-peaked \ions{He}{ii} 4686 line moving in anti-phase, we similarly measure a radial velocity semi-amplitude of \rvcorrHeII. The two peaks in the profile are red- and blue-shifted by \rvshiftHeII.

Finally, the spectra were phase-folded on trial periods of up to 6 hours, but no radial-velocity or intensity variations in the spectral lines were seen on periods other than 5.4 minutes.

\section{Discussion and Conclusions}
\label{sec:conclusion}

\begin{figure}
\begin{center}
\includegraphics[width=84mm]{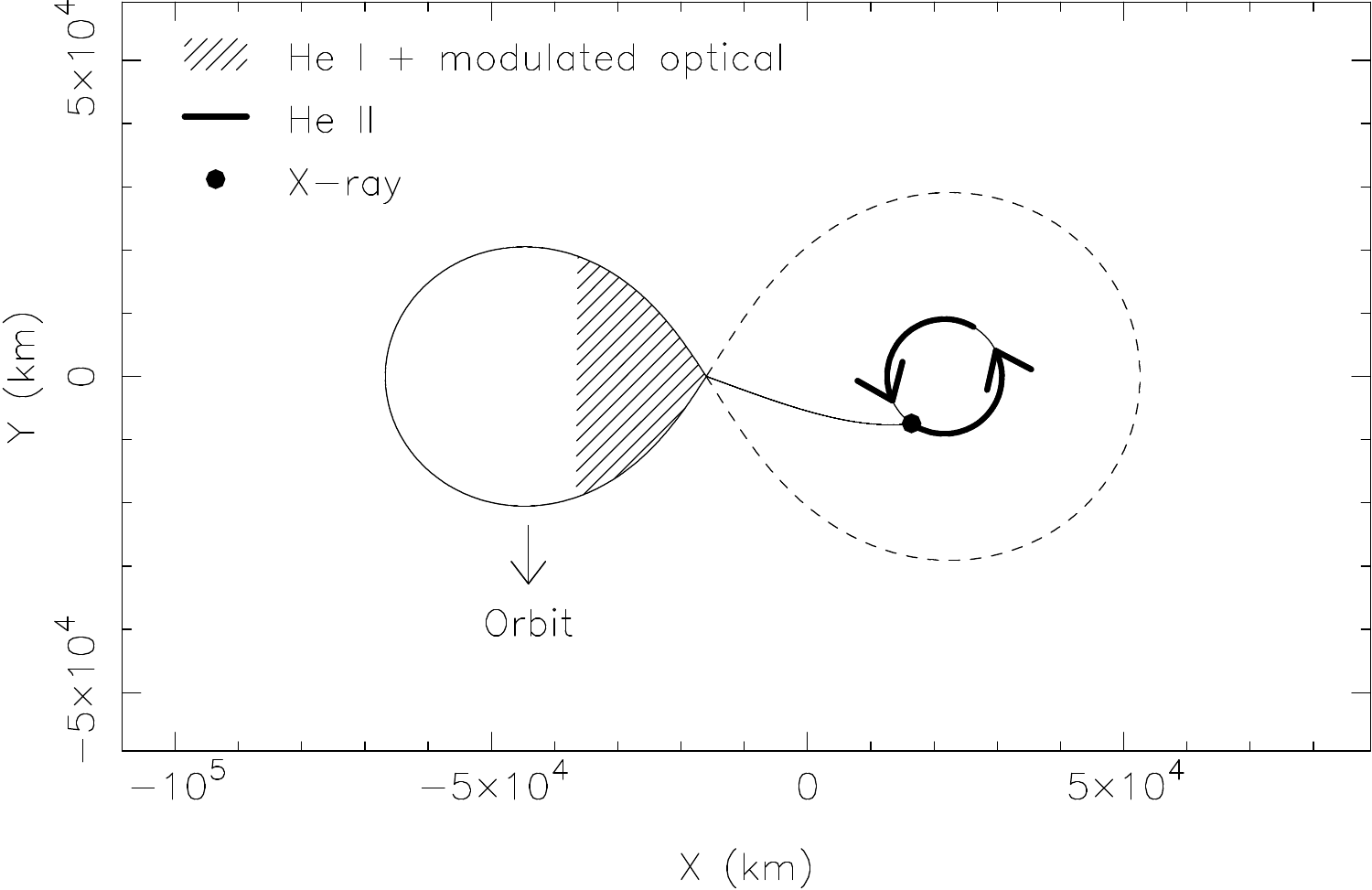}
\caption{`AM CVn' model of \obj, with the donor star on the left, transferring mass to the accretor. The impact of the accretion stream into the accretor causes the X-rays. The irradiated face of the donor star is the source of the \ions{He}{i} 4471 emission line, and the main source of the modulated optical emission. An equatorial belt or disk around the accretor is the source of the \ions{He}{ii} emission lines. Masses $M_2=0.27\,M_\odot$, $M_1=0.55\,M_\odot$ are assumed (see text); the center-of-mass is at the origin.  The orbital phase (from Eq.~(\ref{eq:ephemeris})) would be $\sim$0.35 as viewed from the bottom of this page.}
\label{fig:model}
\end{center}
\end{figure}

\begin{figure}
\begin{center}
\includegraphics[width=84mm]{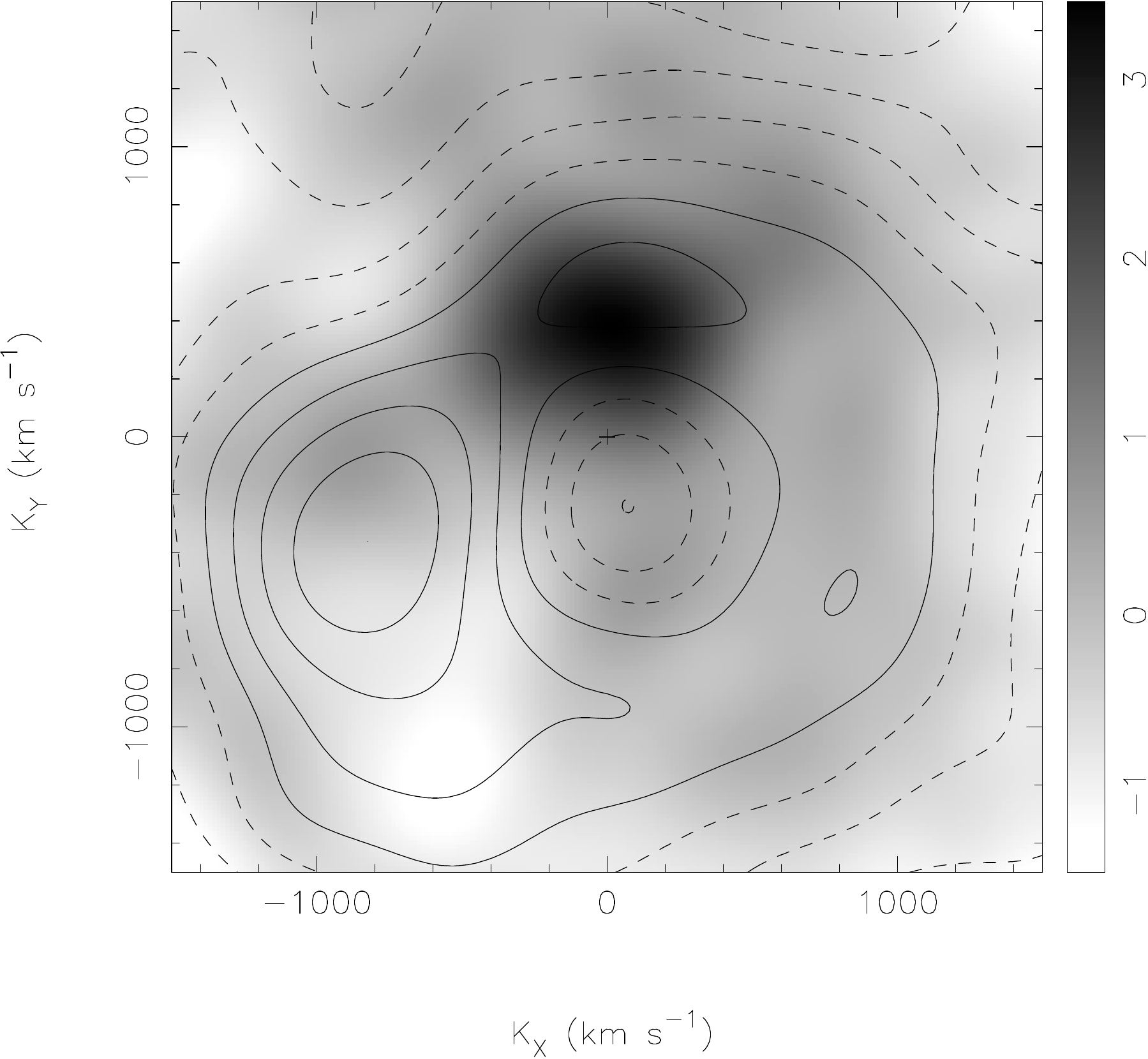}
\caption{Doppler tomograms of \ions{He}{i} 4471 (gray-scale) and \ions{He}{ii} 4686 (contours). The (assumed) irradiation-induced \ions{He}{i} 4471 emission from the secondary star has been aligned with the positive $K_Y$-axis. The gray-scale indicates the flux overdensity as a percentage of the continuum. \ions{He}{ii} 4686 is emitted in a ring centered on the expected location of the primary star, for mass ratios \q. Solid and dashed contours indicate where the \ions{He}{ii} 4686 flux rises above and drops below the average in the map, respectively. A bright spot occurs on the left side of the ring, approximately 110 degrees ahead of the secondary star.}
\label{fig:back}
\end{center}
\end{figure}

\subsection{The Nature of \obj}

The observed radial-velocity variations in the spectral lines strongly suggest that \obj\ has a 5.4-minute orbital period. We favor the semi-detached binary white dwarf (or AM CVn) model, shown in Figure \ref{fig:model}. It is the only model that can account for the spectroscopy presented in this Letter. The IP model predicts an orbital period of hours rather than minutes, and the observed line kinematics do not match predictions by \citet{Nor04} for spectral line variability on the accretor's spin period (if any). We see the broad, fairly constant (and in the case of \ions{He}{ii} 4686, double-peaked) \ions{He}{ii} emission as a clear signature of accretion, which is not predicted by the UI model.

We conclude that \obj\ is by far the shortest-period binary star known. Its direct progenitor was probably a detached binary white dwarf, since evolutionary models with initially non-degenerate donor stars cannot reach such short orbital periods (e.g.\ \citealt{Yun08,Slu05}).

The observed decrease of \obj's 5.4-minute period, considered the main obstacle for the AM CVn model, may be expected naturally if it is a semi-detached binary white dwarf with this orbital period, due to the relatively long mass-transfer turn-on timescales predicted for fairly low-mass helium donors (\citealt{Del06}; Roelofs \& Deloye in preparation).

\subsection{System Parameters}

Based on the radial velocities of the \ions{He}{i} 4471 and \ions{He}{ii} 4686 emission lines moving in anti-phase, we can estimate the ratio of donor to accretor mass $q\equiv M_2/M_1$ in \obj. The \ions{He}{i} 4471 line seems to originate from the irradiated face rather than the exact mass center of the donor star, and so we must estimate a `K-correction' (e.g.\ \citealt{Mun05}) to obtain the donor's actual projected orbital velocity from the observed emission-line velocity. Without a K-correction we get an upper limit of \qhigh. Assuming that the \ions{He}{i} line originates from the inner Lagrange point (the maximum conceivable K-correction) gives a lower limit \qlow. The true value will be in between; assuming a flat probability density distribution between our upper and lower limits gives \q.

Figure~\ref{fig:back} shows a combined Doppler tomogram of the \ions{He}{i} 4471 and \ions{He}{ii} 4686 lines, showing the (putative) irradiated donor star and ring-like emission centered on the (proposed) accreting star. The mass ratio derived above can easily be estimated from this figure. A further interesting feature is the `bright spot' in the \ions{He}{ii} 4686 emission, occurring in the lower left quadrant of the Doppler tomogram. The velocity vectors from the center-of-mass to the donor star and to this \ions{He}{ii} 4686 bright spot make a 110 degree angle approximately. Kinematically, this bright spot matches with a spot approximately on the side of the accreting star as seen from the donor, if the material in the bright spot is moving at roughly half the Keplerian (break-up) velocity near the surface of the accretor (but depending on its mass). An accretion stream impact spot at this location was proposed by \citet{Bar07} to explain the relative phases of the optical and X-ray light-curves. Assuming $M_2=0.13\,M_\odot$, which is the minimum donor star mass corresponding to a Roche-lobe-filling degenerate helium object, an accretor mass $M_1\approx 0.55\,M_\odot$ is required to get an accretion stream impact spot sufficiently on the side of the accretor, as shown by \citet{Bar07}. Our mass ratio \q\ implies that both the donor and the accretor have to be more massive than this to still produce an impact spot sufficiently on the side (but see \citealt{Woo09} for a possible relaxation of this constraint).

The measured rate of change of the orbital period in \obj\ (Eq.~\ref{eq:ephemeris}) also suggests that the donor star has to be significantly more massive than its fully-degenerate value, if one assumes that the rate of change matches the secular rate set by gravitational-wave radiation (neglecting mass transfer). Given $q\approx0.50$, masses $M_2=0.27$, $M_1=0.55$ would be required to yield the observed period derivative. We note that these are quite typical masses for binary white dwarfs according to theoretical models \citep{Nel01}.  Assuming instead $M_2=0.13\,M_\odot$, which is again the minimum mass of a Roche-lobe filling helium donor, an accretor mass $M_1=1.3\,M_\odot$ would be required. This is incompatible with our mass ratio measurement. Note that mass transfer reduces the orbital period change rate relative to the case of a detached binary, so that higher masses would be required; but in practice, it is likely that the mass transfer is not yet sufficiently developed to have a big influence \citep{Del07}. We can conclude that the donor star is probably at least twice as massive as a fully-degenerate helium donor would be. For masses $M_2=0.27\,M_\odot$, $M_1=0.55\,M_\odot$ and assuming that the radial-velocity semi-amplitude of the wobbling \ions{He}{ii} 4686 line represents the projected orbital velocity of the accretor, we obtain an inclination $i\approx38^\circ$ of the orbital plane.

\subsection{\obj\ as a Gravitational Wave Source}

With the system parameters derived in the previous section, we can calculate the gravitational-wave strain amplitude at Earth. The distance to \obj\ is the largest remaining uncertainty; it is probably $\sim$5 kpc based on the expected sizes and measured temperatures of the white dwarf components \citep{Bar07}. This gives a dimensionless gravitational-wave strain amplitude $h\simeq 1.0\times 10^{-22}$. This, together with the design sensitivities of space-borne gravitational wave detectors like the \emph{Laser Interferometer Space Antenna (LISA)}, makes \obj\ one of the easiest detectable sources of gravitational waves known (see \citealt{Nel09}).

\acknowledgments

We thank Susana Barros for helping us with the ephemeris of HM Cnc. GHAR was supported by NWO Rubicon grant 680.50.0610 to G.H.A. Roelofs. AR  acknowledges  support  through  NASA  grant  NNX08AK66G. Based on data obtained at the W.~M. Keck Observatory, which is operated as a scientific partnership among the California Institute of Technology, the University of California and the National Aeronautics and Space Administration. The Observatory was made possible by the generous financial support of the W.M. Keck Foundation.

\end{document}